\documentclass[final,5p,times,twocolumn]{elsarticle}

\usepackage{amssymb}
\usepackage{color}

\journal{Physics Letters B}

\begin{document}

\begin{frontmatter}



\title{Nuclear landscape in covariant density functional theory.}


\author[msu]{A.\ V.\ Afanasjev}
\corref{cor1}
\ead{afansjev@erc.msstate.edu} 

\author[msu]{S.\ E.\ Abgemava}

\author[msu]{D.\ Ray}

\author[tum]{P.\ Ring}

\address[msu]{Department of Physics and Astronomy, Mississippi State
University, MS 39762}

\address[tum]{Fakult\"at f\"ur Physik, Technische Universit\"at M\"unchen,
 D-85748 Garching, Germany}

\cortext[cor1]{Corresponding author}

\begin{abstract}
 The neutron and proton drip lines represent the limits of the nuclear landscape.
While the proton drip line is measured experimentally up to rather high $Z$-values,
the location of the neutron drip line for absolute majority of elements is based 
on theoretical predictions which involve extreme extrapolations. The first ever 
systematic investigation of the location of the proton and neutron drip lines 
in the covariant density functional theory has been performed by employing 
a set of the state-of-the-art parametrizations. Calculated theoretical 
uncertainties in the position of two-neutron drip line are compared with those 
obtained in non-relativistic DFT calculations. Shell effects drastically affect 
the shape of two-neutron drip line. In particular, model uncertainties in the 
definition of two-neutron drip line at $Z\sim 54, N=126$ and $Z\sim 82, N=184$ 
are very small due to the impact of spherical shell closures at $N=126$ and 
184.
\end{abstract}


\end{frontmatter}




  At present, the nuclear masses of approximately 3000 out of
roughly 7000 nuclei expected between nuclear drip lines are known 
\cite{AME2012}. Nuclear existence ends at the drip lines. While 
the proton drip line has been delineated in experiment up to 
protactinium ($Z=91$), the position of the neutron drip line beyond 
$Z=8$ is determined only in model calculations. Different models 
and different parameterizations show rather large variations in 
predictions of the neutron drip line. Moreover, because of experimental 
limitations even in foreseeable future it will be possible to define 
the location of neutron-drip line for the majority of elements only 
in model calculations. In such a situation it is important to estimate 
the errors in the location of the predicted neutron drip line introduced
by the use of the various calculations. In this context we have 
to distinguish  the results and related theoretical 
uncertainties obtained within the same model, but  with different 
parameterizations and the results and uncertainties obtained 
with different models. 

 Theoretical uncertainties(errors) in the prediction of physical 
observables have several sources of origin. Within one class of 
models they are the consequences of specific assumptions and the 
optimization protocols. The differences in the basic assumptions 
of different model classes is another source. They lead to theoretical 
uncertainties which can be revealed only by a systematic comparison
of a variety of models.

 The first attempt to estimate theoretical uncertainties in 
the definition of two-neutron drip line within one class of 
models has been performed within the Skyrme density functional 
theory (SDFT) in Ref.\ \cite{Eet.12} employing the set of 
six parametrizations. These results 
were compared with those obtained in other classes of 
non-relativistic models such as the microscopic-macroscopic 
finite range droplet model (FRDM) \cite{MNMS.95} and the Skyrme 
Hartree-Fock-Bogoliubov (HFB) calculations of Ref.\ \cite{GCP.10} with 
the HFB-21 parametrization. It turns out that the two-neutron drip lines 
of the FRDM and Skyrme-HFB calculations are  located either within the 
SDFT error band or very close to it. Similar calculations exist also 
for non-relativistic DFT models based on the finite range Gogny 
forces D1S \cite{DGLGHPPB.10} and D1M \cite{GHGP.09}.

 The question of theoretical errors in the definition of the neutron 
drip line is still not resolved since the important class of nuclear 
structure models known under name covariant density functional theory 
(CDFT) \cite{Serot1986_ANP16-1,Reinhard1989_RPP52-439,Ring1996_PPNP37-193,
VALR.05,Meng2006_PPNP57-470} has not been applied so far in a reliable way 
to the study of this quantity. Typically, non-relativistic and 
relativistic DFT differ significantly in the prediction of separation 
energies close to the drip lines and, in general, of isovector 
properties far from stability \cite{V.05}. This may lead to neutron 
drip lines which differ substantially from non-relativistic models. 
The goals of the present manuscript are (i) the systematic study of 
two-proton- and two-neutron-drip lines within the relativistic 
Hartree-Bogoliubov (RHB) framework \cite{Kucharek1991_ZPA339-23,
GonzalesLlarena1996_PLB379-13} using several
state-of-the-art CDFT parametrizations, (ii) the estimate of 
theoretical errors in the location of the drip lines within
CDFT framework, and (iii) the comparison of the 
drip lines obtained in relativistic and non-relativistic DFT and 
thus the estimate of global theoretical errors.

\begin{figure*}[ht]
\includegraphics[width=8.5cm,angle=-90]{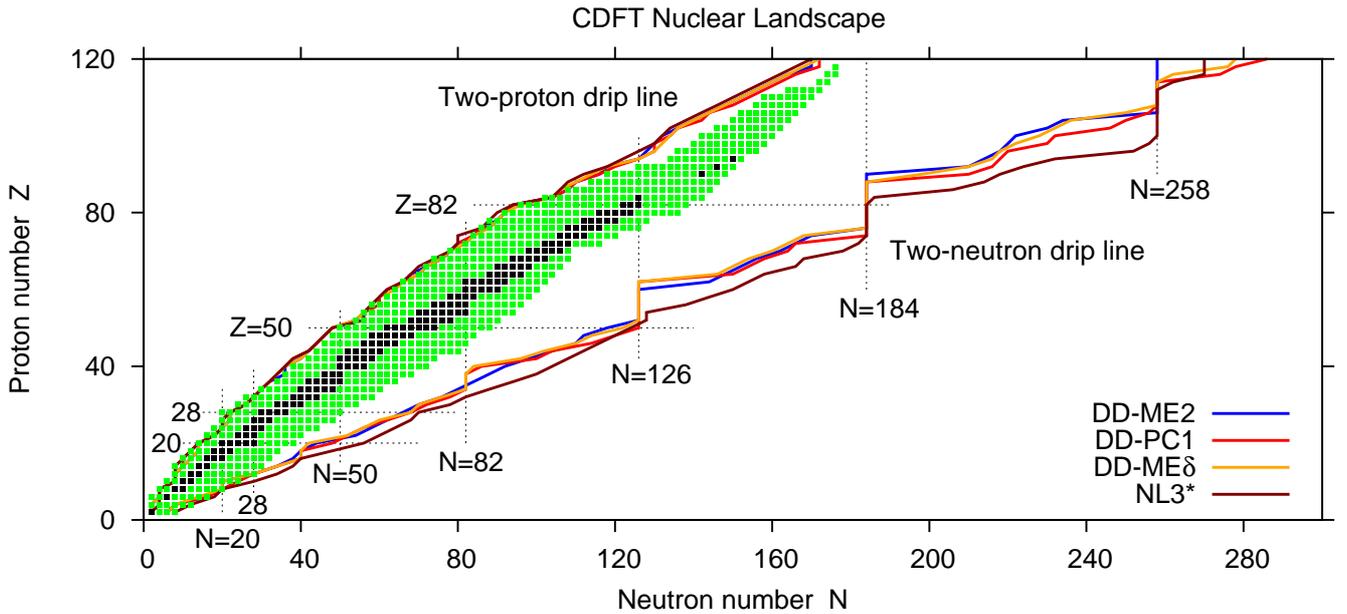}
\caption{The landscape of bound even-even nuclei as
obtained in the CDFT calculations. Experimentally known stable and
radioactive nuclei are shown by black and green squares, respectively.
The experimental data are from Ref.\ \protect\cite{AME2012}. Two-proton  
and two-neutron drip lines calculated with different CEDF are shown by 
the lines of different color.
\label{chart} }
\end{figure*}

 To our knowledge, there were only two previous attempts to study 
the neutron-drip line in the CDFT frawework \cite{HSet.97,GTM.05}. 
However, both of  them employ quite crude approximations to the physics 
of drip line nuclei with a rather limited validity. For example, the 
pairing correlations have been completely ignored in the studies of Ref.\ 
\cite{HSet.97} and the treatment of pairing via BCS approximation in 
Ref.\ \cite{GTM.05} is questionable in the region of drip line since
this approximation does not take into account the continuum properly
and leads to the formation of a neutron gas \cite{DFT.84} in  nuclei 
near neutron-drip line. In addition, these calculations use at most 
14 fermionic shells for the harmonic oscillator basis, which according 
to our study and the one of Ref.\ \cite{RA.11} is not sufficient for a 
correct description of binding energies of actinides and superheavy nuclei 
and the nuclei in the vicinity of neutron-drip line. The RHB framework with 
a finite range pairing force is a proper tool for that purpose. It has been 
applied very successfully with the parameter set NL3 \cite{LVR.01,LVR.04} 
and the parameter set DD-PC1~\cite{Ferreira2011_PLB701-508} at the proton 
drip line and it has the proper coupling to the continuum at the neutron 
drip line.

 In the present manuscript, the RHB framework is used for a systematic 
studies of ground state properties of all even-even nuclei from the proton- 
to neutron drip line. The separable version \cite{TMR.09,Tian2009_PRC80-024313} 
of the finite range Brink-Booker part of the Gogny D1S force is used in the 
particle-particle channel; its strength variation across the nuclear chart 
is defined by means of the fit of rotational moments of inertia calculated 
in the cranked RHB framework to experimental data via the procedure of Ref.\ 
\cite{AA.13}. The need for such $A$-dependent variation of the strength of 
the Brink-Booker part of the Gogny D1S force in the CDFT application has 
recently been discussed in Refs.\ \cite{AA.13,WSDL.13}. As the absolute 
majority of nuclei are known to be axially and reflection symmetric in their 
ground states, we consider only axial and parity-conserving
intrinsic states and solve the RHB-equations in an axially deformed oscillator 
basis \cite{Gambhir1990_APNY198-132,Ring1997_CPC105-77}. The truncation of the 
basis is performed in such a way that all states belonging to the shells up to
$N_F = 20$ fermionic shells and $N_B = 20$ bosonic shells are taken into 
account. This provides sufficient numerical accuracy. As the absolute
majority of nuclei are known to be axially and reflection symmetric in 
their ground states, we consider only axial and parity-conserving 
intrinsic states. For each nucleus the potential energy curve in large 
deformation range from $\beta_2=-0.4$ up to $\beta_2=1.0$ is obtained 
by means of constraint on the quadrupole moment $Q_{20}$. Then, the 
correct ground state configuration and its energy are defined; this 
procedure is especially important for the cases of shape coexistence.

In axial reflection-symmetric calculations  for superheavy nuclei with 
$Z\geq 100$, the superdeformed minimum is frequently lower in energy 
than the normal deformed one \cite{AAR.12}. As long as triaxial and
octupole deformations are not included, this minimum is stabilized
by the presence of an outer fission barrier. Including such deformations,
however, it often turns out that this minimum either disappears
or becomes a saddle point, unstable against fission~\cite{AAR.12}.
Since these deformations are not included in the present calculations, 
we restrict our consideration to spherical or normal-deformed ground 
states in the $Z\geq 100$ nuclei. This also facilitates the comparison 
with non-relativistic results which favor such ground states for these 
nuclei.

 
 Three existing classes of covariant density functional models 
are used throughout this paper: the nonlinear meson-nucleon
coupling model (NL), the density-dependent meson-exchange
model (DD-ME), and a density-dependent point coupling model 
(DD-PC); see their comparison in Ref.\ \cite{AAR.12}. The main 
differences among them lay in the treatment of the range of 
the interaction,  the mesons, and the density dependence. 
The interaction in the first two classes has a finite range, 
while  the third class uses a zero-range interaction
with one additional gradient term in the scalar-isoscalar
channel. The mesons are absent in the density-dependent point
coupling model. The density dependence is explicit in the last
two models, while it shows up via the nonlinear meson-couplings
in the first case.

  Each of these model classes is represented here by the energy 
density functional (EDF) that is considered to be the state-of-the-art.  
The NL model is represented here by the NL3* \cite{NL3*} EDF
which has the smallest number of parameters amongst considered
EDF fitted to data.
The DD-ME model is 
represented by the DD-ME2 \cite{DD-ME2} and the DD-ME$\delta$ 
\cite{DD-MEdelta} EDFs. The DD-ME$\delta$ EDF differs from others 
by the inclusion of the $\delta$-meson, which leads to different 
proton and neutron effective masses. In addition, the parameters 
of the DD-ME$\delta$ EDF are largely based on microscopic 
{\it ab initio} calculations in nuclear matter; 
only four of its parameters are fitted to finite nuclei. On the
contrary, all parameters of other EDF were adjusted to experimental
data based on the properties of finite nuclei. The DD-PC 
model is represented by the DD-PC1 \cite{DD-PC1} EDF. In contrast 
to the other functionals, which are fitted to spherical nuclei, 
this EDF is fitted to a large set of deformed nuclei.


\begin{figure*}[ht]
\includegraphics[width=18cm,angle=0]{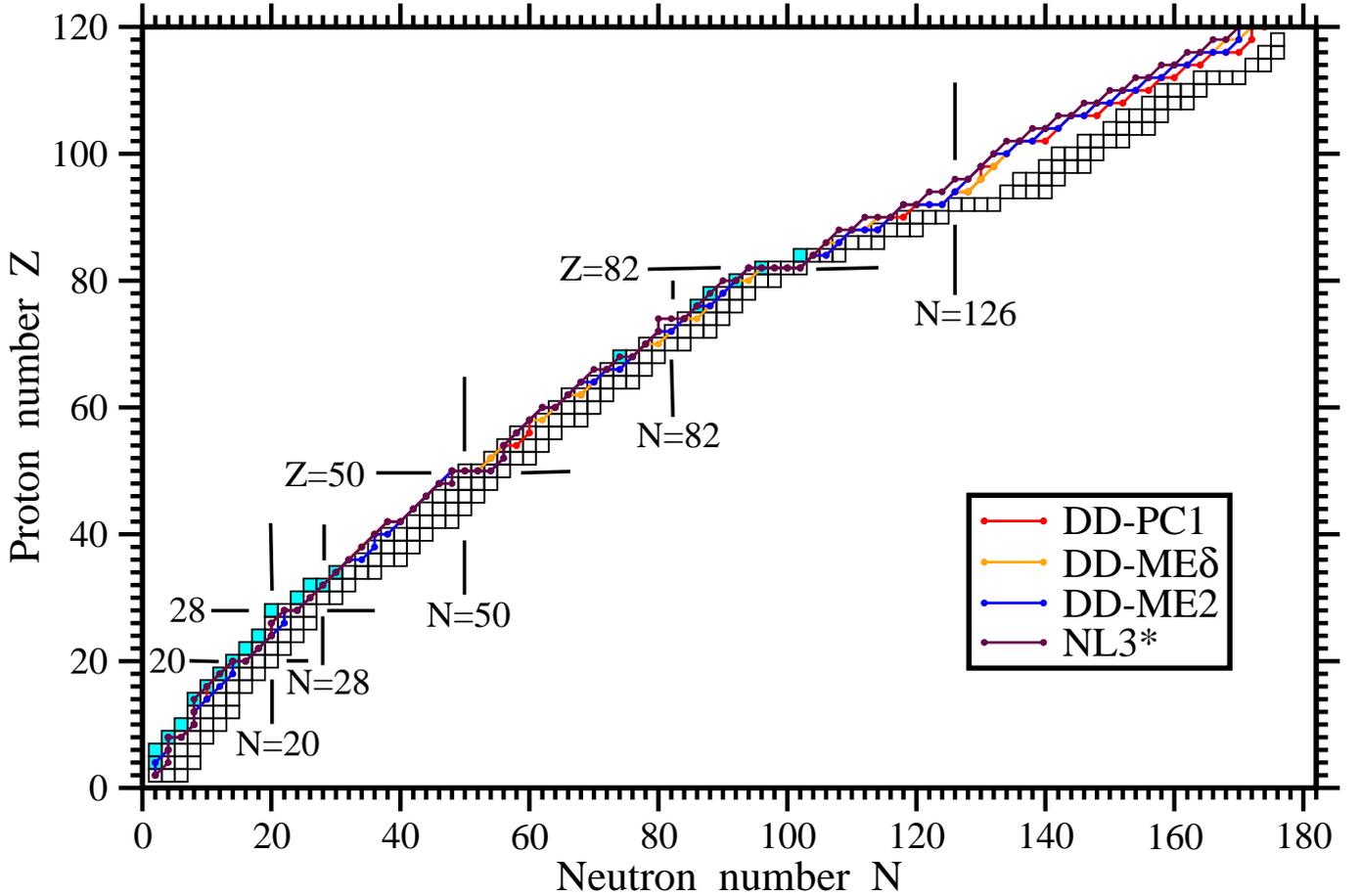}
\centering
\caption{Calculated versus experimental two-proton 
drip lines. For each isotope chain, the four experimentally known most 
proton-rich nuclei are shown by squares. Cyan shading of the squares is used for the 
nuclei located  beyond  the two-proton drip line ($S_{2p}<0)$. The experimental 
data are from Ref.\ \protect\cite{AME2012}.  The borderline between shaded 
and open squares delineates known two-proton drip line. Only in the case of 
the $Z=4,6,8,80,82$ and 84 isotope chains, the location of two-proton drip 
line is firmly established since the masses of the nuclei 
on both sides of the drip line are directly and accurately measured. 
The two-proton drip line is only tentatively delineated for the 
$Z=10,14-34,68,76$ and 78 chains since the 
masses of beyond the drip line nuclei are only estimated in Ref.\ 
\protect\cite{AME2012}. The lines with small symbols show the calculated 
two-proton drip lines which go along the last two-proton bound nuclei. }
\label{chart-odd}
\end{figure*}

 Fig.\ \ref{chart} shows the nuclear landscape as obtained with these 
CDFT parametrizations. The particle stability (and, as a consequence, 
a drip line) of a nuclide is specified by its separation energy, 
namely, the amount of energy needed to remove particle(s). Since 
our investigation is restricted to even-even nuclei, we consider 
two-neutron $S_{2n}=B(Z,N-2)-B(Z,N)$ and two-proton 
$S_{2p}=B(Z-2,N)-B(Z,N)$ separation energies. Here $B(Z,N)$ stands 
for the binding energy of a nucleus with $Z$ protons and $N$ neutrons. 
If the separation energy is positive, the nucleus is stable against 
two-nucleon emission; conversely, if the separation energy is 
negative, the nucleus is unstable. Thus, two-neutron and two-proton 
drip lines are reached when $S_{2n}\leq 0$ and $S_{2p}\leq 0$, 
respectively.

\begin{table}[ht] 
\caption{The rms-deviations $\Delta E_{rms}$, $\Delta (S_{2n})_{rms}$
($\Delta (S_{2p})_{rms}$) between calculated and experimental binding 
energies $E$ and two-neutron(-proton) separation energies $S_{2n}$
($S_{2p}$), respectively. They are given in MeV for indicated CDFT 
parametrizations with respect of ``measured'' and ``measured+estimated'' 
sets of experimental masses.}
\begin{tabular}{|c|c|c|c|c|} \hline
EDF   & measured  & \multicolumn{3}{|c|}{measured+estimated}   \\ \hline
  & $\Delta E_{rms}$  & $\Delta E_{rms}$ & $\Delta (S_{2n})_{rms}$ & $\Delta (S_{2p})_{rms}$ \\ \hline
   NL3*            &    2.97  &  3.01 & 1.21 & 1.28 \\
   DD-ME2          &    2.42  &  2.48 & 1.09 & 0.99 \\
   DD-ME$\delta$   &    2.31  &  2.42 & 1.11 & 1.11 \\
   DD-PC1          &    2.02  &  2.17 & 1.25 & 1.13 \\ \hline
\end{tabular}
\label{deviat}
\end{table}


  The accuracy of the description of separation energies depend 
on the accuracy of the description of mass differences. The 
global RHB calculations of masses with employed parametrizations 
lead to the rms-deviations $\Delta E_{rms}$ between calculated 
and experimental binding energies which are listed in Table 
\ref{deviat}. The detailed results of these calculations will 
be presented in a forthcoming manuscript \cite{AARR.13}. The masses 
given in the AME2012 mass evaluation \cite{AME2012} can be 
separated into two groups; one represents nuclei with masses 
defined only from experimental data, the other contains nuclei
with masses depending in addition on either interpolation or
extrapolation procedures. For simplicity, we call the masses of 
the nuclei in the first and second groups as measured and estimated. 
There are 640 measured and 195 estimated masses of even-even nuclei 
in the AME2012 mass evaluation.  One can see  in Table \ref{deviat} 
that the addition of estimated masses leads only to a slight 
decrease of the accuracy of the description of experimental 
data. Two-neutron $S_{2n}$ and two-proton $S_{2p}$ separation 
energies are described with typical accuracy of 1 MeV (Table 
\ref{deviat}). One can see that not always the parametrization
which provides the best description of masses gives the best
description of two-particle separation energies. This is 
because the separation energies are related to the derivatives
of binding energies with respect of particle number.


  Fig.\ \ref{chart-odd} shows that theoretical uncertainties (i.\ e. the
spread of the predictions due to different EDF) are rather small for 
two-proton drip line. In addition, the results of the calculations are 
very close to experimental data. This is because the proton-drip line 
lies close to the valley of stability, so that extrapolation errors 
towards it are small. Another reason is the fact the Coulomb barrier 
provides a rather steep potential reducing considerably the coupling 
to the proton continuum. This leads to a relatively low density of
the single-particle states in the vicinity of the Fermi level.


  The situation is different for the two-neutron drip line. In the 
majority of the cases, the theoretical uncertainties in the location 
of this line are much larger than for the two-proton drip one and they 
are generally increasing with the increase of mass number. This 
is commonly attributed to poorly known isovector properties 
of EDF \cite{Eet.12}. Although this factor contributes, such an
explanation is somewhat oversimplified from our point of view. 
That is because for some combinations of $Z$ and $N$ there is basically 
no (or very little) dependence of the predictions for the location 
of the two-neutron drip line on the CDFT parametrization. Such a weak
(or vanishing) dependence is especially pronounced at spherical neutron 
shell closures with $N=126,184$ and 258 around proton numbers $Z=54,80$ 
and 110. It is interesting that the impact of shell structure at 
these particle numbers on the shape of the two-neutron drip line is more 
pronounced than that for the  two-proton drip line due to $Z=50$ and 82 
proton shell  gaps.

  However, moving away from these spherical shell closures the 
spread of theoretical predictions for the two-neutron drip line 
increases. This move also induces the deformation in the nuclei.  
Thus, there is a close correlation between the nuclear deformation
at the neutron-drip line and the uncertainties in the prediction 
of neutron-drip line; the regions of large uncertainties corresponds to 
transitional and deformed nuclei. This is caused by the underlying
densities of the single-particle states. The spherical nuclei under 
discussion are characterized by large shell gaps and a clustering 
of highly degenerate single-particle states around them. Deformation
removes this high degeneracy of single-particle states and leads to a 
more equal distribution of the single-particle states with energy. 
Moreover, the density of bound neutron single-particle states close 
to the neutron continuum is substantially larger than that on the 
proton-drip line. As a consequence, inevitable inaccuracies in the 
DFT description of the deformed single-particle state energies which 
are present even in the valley of beta-stability \cite{AS.11} will 
lead to larger uncertainties in the predictions of the neutron-drip
line.

\begin{figure}[ht]
\includegraphics[width=8cm,angle=0]{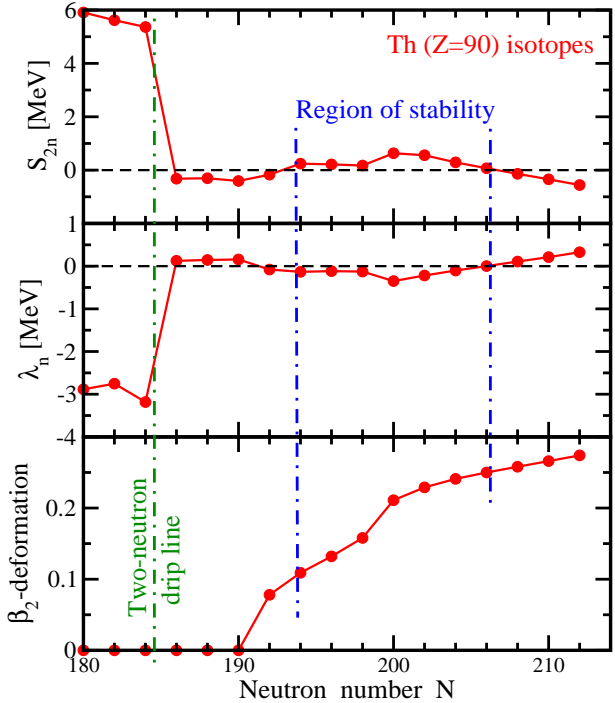}
\caption{Two-neutron separation energies $S_{2n}$,
neutron chemical potentials $\lambda_n$, and quadrupole deformations
$\beta_2$ of the Th$(Z=90)$ isotopes obtained in the RHB(DD-ME2)
calculations.}
\label{reemer}
\end{figure}


 For some isotope chains, 
there are regions of two-neutron stability (not shown in Fig.\ 
\ref{chart}) at neutron numbers beyond the primary two-neutron 
drip line. The physical mechanism behind the appearance of 
these regions is illustrated in Fig.\ \ref{reemer} on the 
example of the Th isotope chain. Two-neutron separation 
energies $S_{2n}$ and the neutron chemical potential $\lambda_{2n}$ 
are positive and negative in two-neutron bound nuclei ($N\leq 184$), 
respectively. The $S_{2n}$ and $\lambda_{2n}$ values become negative and 
positive for two-neutron unbound nuclei ($186\leq N\leq 192$),
respectively. A further increase of the neutron number triggers an
increase of quadrupole deformation $\beta_2$ leading to a 
lowering of the neutron chemical potential $\lambda_n$ which 
again becomes negative. As a consequence, two-neutron binding 
reappears ($S_{2n}>0$) at $N=194-206$. Further increase of
$N$ beyond 206 leads to two-neutron unbound nuclei. The 
appearance of these regions, however, strongly depends on 
the CDFT parametrization. For example, such regions exist 
at $(Z=62,N=132-146)$, $(Z=88,N=194-206)$ for DD-PC1, at 
$(Z=74,N=176-184)$, $(Z=90,N=194-206)$ for DD-ME2 and at 
$(Z=62,N=132-142)$, $(Z=74,N=178-184)$ and $(Z=90,N=204-206)$ for
DD-ME$\delta$. However, the regions of stability beyond the 
primary drip line are absent in the RHB(NL3*) calculations.

A similar reappearance of two-neutron binding with increasing
neutron number beyond primary two-neutron drip line exists 
also in many SDFT parametrizations \cite{Eet.12}. 
Both in CDFT and SDFT, the regions of two-neutron 
binding reappearance represent the peninsulas emerging from 
the nuclear mainland. Ref.\ \cite{Eet.12} suggested
that such behavior is due to the presence of shell effects 
at neutron closures that tend to lower binding energy along 
the localized bands of stability. This is certainly true
in some cases. However, our analysis 
presented above suggests that local changes of the shell 
structure induced by deformation changes play also an important
role. Similar to the CDFT(NL3*) results, there are also some 
Skyrme EDF which do not show the reappearance of two-neutron 
binding \cite{Eet.13}.

\begin{figure*}[ht]
\includegraphics[width=7.8cm,angle=-90]{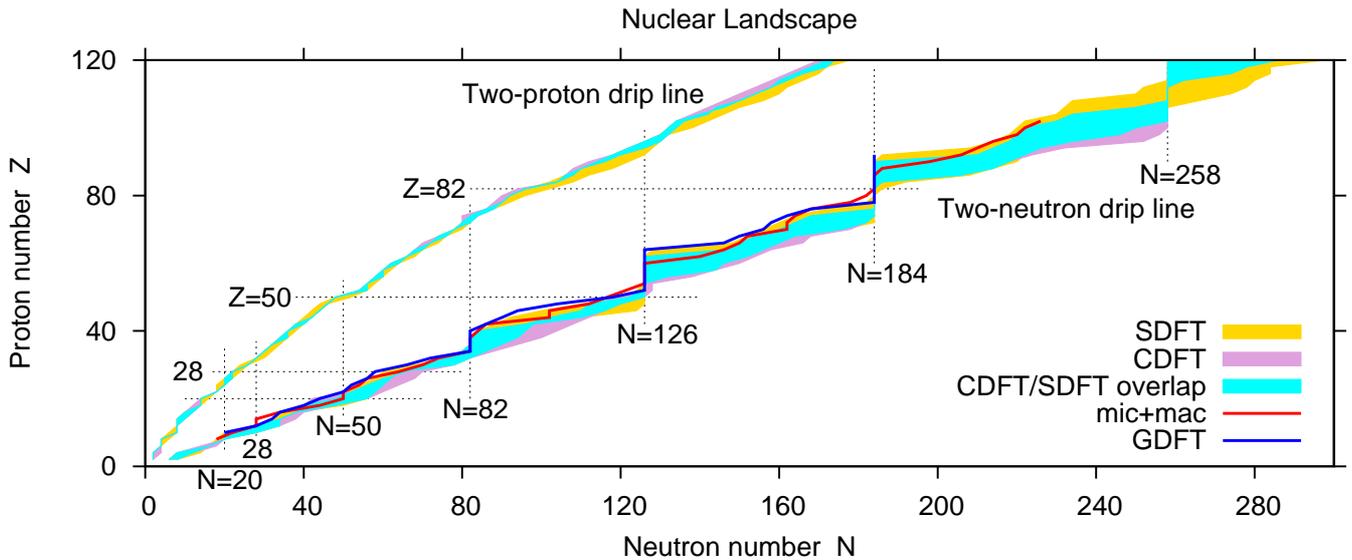}
\caption{The comparison of the uncertainties in the definition of 
two-proton and two-neutron drip lines obtained in CDFT and SDFT. 
The shaded areas are defined by the extremes of the predictions of 
the corresponding drip lines obtained with different parametrizations. 
The blue shaded area shows the area where the CDFT and SDFT results 
overlap. Non-overlapping regions are shown 
by dark yellow and plum colors for SDFT and CDFT, respectively. 
The results of the SDFT calculations are taken from the supplement 
to Ref.\ \protect\cite{Eet.12}. The two-neutron drip lines obtained by 
microscopic+macroscopic (FRDM \protect\cite{MNMS.95}) and Gogny D1S DFT 
\protect\cite{DGLGHPPB.10} calculations  are shown by red and blue lines, 
respectively.}
\label{chart-shade}
\end{figure*}

 It is interesting to compare theoretical CDFT uncertainties 
in the definition of the two-proton and two-neutron drip lines 
with the ones obtained in non-relativistic calculations. Fig.\ 
\ref{chart-shade} presents such a comparison. We use so-called
'2012 Benchmark uncertainties'' \cite{Eet.13} obtained in Ref.\ 
\cite{Eet.12} for Skyrme DFT employing six parametrizations;
these uncertainties are shown by the combination of yellow and
blue shaded areas in Fig.\ \ref{chart-shade}. The CDFT
uncertainties are represented by the combination of the 
plum and blue shaded areas. One can see that the CDFT and SDFT 
uncertainties in the definition of two-proton drip line are 
small; they tightly overlap at $Z\leq 70$ while for higher 
$Z$ the CDFT uncertainties are shifted slightly towards neutron
deficient nuclei as compared with the SDFT ones. The uncertainties 
for two-neutron drip line are larger but still they are similar 
in two models in many regions. In particular, the two-neutron drip 
line at $Z\sim 54, N=126$ and $Z\sim 82, N=184$ is well defined not
only in the CDFT and SDFT calculations, but also in the mic+mac 
(FRDLM) and Gogny D1S calculations. This uniqueness is due to 
corresponding well pronounced spherical shell closures in the 
model calculations.

  The predictions of the DD-ME2, DD-ME$\delta$ and DD-PC1 parametrizations 
are close to each other (Fig.\ \ref{chart}) and are within the '2012 
Benchmark uncertainties'.  The NL3* parametrization typically 
shifts the two-neutron drip line to a higher $N$-value exceeding 
in some regions '2012 Benchmark uncertainties'. However, the same 
is true for recently fitted Skyrme TOV-min parametrization \cite{Eet.13}, 
the two-neutron drip line of which is very similar to the one obtained 
in the RHB(NL3*) calculations.

   The biggest difference between CDFT and Skyrme DFT calculations
appears at $N=258,Z\sim 110$ (see Fig.\ \ref{chart-shade}) where 
the two-neutron drip line is uniquely defined in the CDFT calculations 
due to large spherical gap at $N=258$. This gap is also present
in many Skyrme EDF but it does not prevent a significant spread
of Skyrme DFT predictions for the two-neutron drip line in this region. 
This again underlines the importance of shell structure in the 
predictions of the details of the two-neutron drip line. A similar 
difference between CDFT and SDFT exists also in superheavy nuclei
with $Z\approx 120-126, N\approx 172-184$ where different 
centers of islands of stability are predicted by these models 
\cite{BRRMG.98,AKF.03}. These results are contrary to the fact 
that both models generally agree for lighter $Z\leq 100$ nuclei.

  The DD-* CEDF predict two-neutron drip line at lower $N$ as 
compared with the NL3* one (see Fig.\ \ref{chart}). It is tempting 
to associate this feature with different symmetry energies $J$ 
($J\sim 32$ MeV for DD* and $J\sim 39$ MeV for NL3*). However, a 
detailed analysis of 14 two-neutron drip lines obtained in relativistic 
and non-relativistic calculations does not reveal clear correlations 
between the location of two-neutron drip line and the nuclear matter 
properties of the employed force.


  In conclusion, a detailed analysis of two-neutron drip lines in
covariant and non-relativistic DFT has been performed. These results
clearly indicate that the shell structure is not washed near or at
two-neutron drip line. In particular, model uncertainties in the
definition of two-neutron drip line at $Z\sim 54, N=126$ and 
$Z\sim 82, N=184$ are very small due to the impact of
spherical shell closures at $N=126$ and 184. The largest difference 
between covariant
and Skyrme DFT exist in superheavy nuclei, where the first model
(contrary to second) predicts significant impact of the $N=258$
spherical shell closure. The spread of theoretical predictions
grows up on moving away from these spherical closures. The 
development of deformation causes it. Both poorly known isovector 
properties of the forces and inevitable inaccurcies in the description
of deformed single-particle states in the DFT framework contribute
to that. The number of particle-bound even-even $Z\leq 120$ nuclei 
is 2040, 2050, 2057 and 2216 in the DD-PC1, DD-ME2, DD-ME$\delta$ and
NL3* parametrizations, respectively. This is close to the numbers
obtained in SDFT. Thus, our calculations support the estimate of 
Ref.\ \cite{Eet.12} that around 7000 different (including odd- and
odd-odd ones) nuclides have to exist.

The authors would like to thank J.\ Erler for valuable discussions.
This work has been supported by the U.S. Department of Energy under 
the grant DE-FG02-07ER41459 and by the DFG cluster of excellence 
\textquotedblleft Origin and Structure of the Universe 
\textquotedblright\ (www.universe-cluster.de). This research was 
also supported by an allocation of advanced computing resources 
provided by the National Science Foundation. The computations were 
partially performed on Kraken at the National Institute for 
Computational Sciences (http://www.nics.tennessee.edu/).


\end{document}